\documentclass[a4paper,usenatbib]{mnras}

\usepackage{mathptmx}
\usepackage[T1]{fontenc}
\usepackage{ae,aecompl}

\usepackage{graphicx}   % Including figure files
\usepackage{amsmath}    % Advanced maths commands
\usepackage{amssymb}    % Extra maths symbols

\title[Testing the isotropy of Universe with SNe Ia]{Testing the isotropy of the Universe with type Ia supernovae in a model-independent way}

\author[Wang \& Wang]{
    Yu-Yang Wang$^1$ and F. Y. Wang$^{1,2}$\thanks{E-mail:
        fayinwang@nju.edu.cn}
    \\
    $^{1}$School of Astronomy and Space Science, Nanjing University, Nanjing 210093, China\\
    $^{2}$Key Laboratory of Modern Astronomy and Astrophysics (Nanjing University), Ministry of Education, Nanjing 210093, China\\
}

\newcommand{\ud}{\mathrm{d}}

\begin{document}
    \label{firstpage}
    \maketitle

\begin{abstract}
In this paper, we study an anisotropic universe model with Bianchi-I
metric using Joint Light-curve Analysis (JLA) sample of type Ia
supernovae (SNe Ia). Because light-curve parameters of SNe Ia vary
with different cosmological models and SNe Ia samples, we fit the
SNe Ia light-curve parameters and cosmological parameters
simultaneously employing Markov Chain Monte Carlo method. Therefore,
the results on the amount of deviation from isotropy of the dark
energy equation of state ($\delta$), and the level of anisotropy of
the large-scale geometry ($\Sigma_0$) at present, are totally
model-independent. The constraints on the skewness and
cosmic shear are $-0.101<\delta<0.071$ and $-0.007<\Sigma_0<0.008$.
This result is consistent with a standard isotropic universe
($\delta=\Sigma_0=0$). However, a moderate level of anisotropy in
the geometry of the Universe and the equation of state of dark
energy, is allowed. Besides, there is no obvious evidence for a
preferred direction of anisotropic axis in this model.
\end{abstract}


    %%%%%%%%%%%%%%%%%%%%%%%%%%%%%%%%%%%%%%%%%%%%%%%%%%

    %%%%%%%%%%%%%%%%% BODY OF PAPER %%%%%%%%%%%%%%%%%%

\section{Introduction}\label{sec:introduction}
Astronomical observations revealed that our Universe is undergoing
an accelerating expansion
(\citealt{Riess98};\citealt{Perlmutter99}), which is one of the most
surprising astronomical discoveries in recent years. Accelerating
expansion implies that the universe is dominated by an unknown form
of energy called `dark energy' with negative pressure, or that
Einstein's theory of gravity fails on cosmological scales and
requires some modifications.

The standard $\Lambda \mathrm{CDM}$ model is established based on
the cosmological principle and parametrization of the Big Bang
cosmological model. It depicts a homogeneous and isotropic universe
on large scales with approximately 30\% matter (including baryonic
matter and dark matter) and 70\% dark energy at present time,
which is consistent with vast majority of several precise
astronomical observations, including cosmic microwave background
(CMB) power spectrum (\citealt{Komatsu11}; \citealt{Planck15}) and
baryon acoustic oscillations (\citealt{Eisenstein05}).

However, the standard cosmological model is challenged by a few
puzzling cosmological observations (\citealt{Perivolaropoulos14})
which may require modifications. Evidences for cosmology anisotropy
have been obtained by the power asymmetry of CMB perturbation maps
(\citealt{Eriksen07}; \citealt{Hoftuft09}; \citealt{Paci10};
\citealt{Mariano13}; \citealt{Zhao15}), the large scale velocity flows
(\citealt{Kashlinsky08}; \citealt{Watkins09};
\citealt{Kashlinsky10}; \citealt{Lavaux10}; \citealt{Feldman10}),
anisotropy in accelerating expansion rate (\citealt{Antoniou10};
\citealt{Mariano12}; \citealt{Yang14}; \citealt{Wang14}), spatial
dependence of the value of the fine structure constant $\alpha$
(\citealt{Webb11}; \citealt{Moss11}; \citealt{King12};
\citealt{Mariano12}; \citealt{Pinho16}) and so on. These puzzles are
in favor of preferred cosmological directions, which seem to violate
the cosmological principle. The so-called `cosmic anomalies'
(\citealt{Perivolaropoulos14}) may either be simply large
statistical fluctuations or have some physical origins, which could
be either geometric or energy-related
(\citealt{Perivolaropoulos14}).

Here, we focus on an anisotropic universe model that has a
plane-symmetric Bianchi-I metric (\citealt{Taub51};
\citealt{Schucker14}), namely ellipsoidal universe
(\citealt{Campanelli06}). The ellipsoidal universe model was first
proposed in \cite{Campanelli06} to solve the CMB quadrupole problem
by assuming a plane-symmetric universe with an eccentricity of order
$10^{-2}$ at decoupling. \cite{Campanelli07} discussed that the
anisotropic expansion can be generated by cosmological magnetic
fields, cosmic domain walls or cosmic strings. The cosmic shear
$\Sigma_0$ and skewness $\delta$ are introduced
(\citealt{Campanelli11a}) to describe the anisotropy level of cosmic
geometry and dark energy fluids, respectively. \cite{Campanelli11c}
analysed Union and Union2 compilation and concluded that an
isotropic universe is consistent with SNe Ia data. However, their
analysis directly used $\mu_{\mathrm{obs}}$ and $\sigma$ obtained in
$\Lambda$CDM model (\citealt{Amanullah10}). The results are
model-dependent, because light-curve parameters change with
different universe models and SNe Ia sample. \cite{Schucker14}
fitted the Bianchi I metric to the Hubble diagram of SNe Ia.

Therefore, we improve the previous research by fitting the SNe Ia
light-curve parameters and cosmological parameters simultaneously.
This paper is organized as follows. In the next section, we
introduce the ellipsoidal universe model and derive the
magnitude-redshift relation. In section 3, we use SNe Ia data of JLA
sample to constrain all the free parameters simultaneously,
including light-curve and cosmological parameters. The fitting
results are shown in section 4. Conclusion and discussions are given
in section 5.

\section{Ellipsoidal universe model}
    \label{sec:ellipsoidal universe model}
The Bianchi type I cosmological model is extensively discussed in
\cite{Campanelli06} and \cite{Campanelli11a}. In this section, we
briefly introduce this model. The Bianchi-I metric
(\citealt{Taub51}; \citealt{Schucker14}) with planar symmetry
(\citealt{Campanelli06}; \citeyear{Campanelli07};
\citealt{Campanelli11a}; \citeyear{Campanelli11b};
\citeyear{Campanelli11c}) is described by Taub line element
(\citealt{Campanelli11a}; \citeyear{Campanelli11b})
    \begin{equation}
    \ud \tau^2 = \ud t^2 - a(t)^2(\ud x^2 + \ud y^2) - b(t)^2\ud z^2, \qquad a(t), b(t) > 0.
    \end{equation}
where $a(t)$ and $b(t)$ are the scale factors which can be
normalized as $a(t_0) = b(t_0) = 1 $ at the present time $t_0$.

According to the plane-symmetric metric, the `mean Hubble parameter'
$H$ can be defined as (\citealt{Campanelli11b};
\citeyear{Campanelli11c})
    \begin{equation}
    H \equiv \frac{\dot{A}}{A}\, ,
    \end{equation}
where $A\equiv(a^2b)^{\frac{1}{3}}$ is the `mean expansion
parameter' (\citealt{Campanelli11b}).

In order to measure the level of anisotropy, we define cosmic shear
$\Sigma$ and skewness $\delta$ (\citealt{Campanelli11a};
\citeyear{Campanelli11b}; \citeyear{Campanelli11c}) as
    \begin{equation}
    \Sigma \equiv \frac{H_a - H}{H}\,, \qquad \mathrm{and} \qquad \delta \equiv w_{\parallel} - w_{\perp}\,,
    \end{equation}
respectively. In the above equation, $H_a = \dfrac{\dot{a}}{a}$
represents the Hubble parameter in the symmetry plane.
$w_{\parallel}$ and $w_{\perp}$ are parameters of state equation
parallel and perpendicular to the symmetry plane respectively.
Besides, the mean parameter of state equation is defined as
    \begin{equation}
    w \equiv \frac{2w_\parallel + w_\perp}{3}\,.
    \end{equation}
    \subsection{Anisotropy axis}
In the Galactic coordinate reference, the direction cosine of the
symmetry axis is
    \begin{equation}
    \hat{n}_A = (\cos b_A\cos l_A, \cos b_A\sin l_A, \sin b_A)\,.
    \end{equation}
For an arbitrary direction
    \begin{equation}
    \hat{n} = (\cos b\cos l, \cos b\sin l, \sin b),
    \end{equation}
the angle $\theta$ between $\hat{n}$ and $\hat{n}_A$ is
    \begin{equation}
    \cos\theta \equiv \hat{n}\cdot\hat{n}_A\,.
    \end{equation}
    \subsection{Redshift-distance relation}
We introduce the `eccentricity' $e$ as
    \begin{equation}
    e^2 \equiv 1- \frac{b^2}{a^2}\,.
    \end{equation}
In galactic coordinates system, the luminosity distance is given by
(\citealt{Campanelli11c})
    \begin{equation}
    d_L(z,\theta) = \frac{c(1+z)}{H_0}\int_{A(z)}^{1}\frac{(1-e^2)^\frac{1}{6}}{(1-e^2\cos^2\theta)^{\frac{1}{2}}}\frac{\ud A}{A^2\bar{H}}\,,
    \end{equation}
where $\bar{H}$ is the Hubble constant normalized to its actual at
$t_0$. Both $\bar{H}$ and $e$ are functions of $A$.

\section{JLA sample and MCMC fitting}
    \label{sec:jla}
The Joint Light-curve Analysis (JLA) sample (\citealt{Betoule14}) is
based on \cite{Conley11} compilation. It includes three-season data
from SDSS-II ($0.05<z<0.4$), three-year data from SNLS ($0.2<z<1$),
HST data ($0.8<z<1.4$) and several low-redshift samples ($z<0.1$)
such as Cal\'{a}n/Tololo Survey and Carnegie Supernova Project. The
JLA sample totals 740 spectroscopically confirmed SNe Ia with
high-quality light curves.

\subsection{Angular position}
    \label{sec:angular position} % used for referring to this section from elsewhere
In order to determine the anisotropy axis in the galactic coordinate
system, we need galactic latitude and longitude $(l, b)$ for each
supernova. For transformations from equatorial to Galactic
coordinates, we have
    \begin{equation}
    \sin b = \sin\delta_{\mathrm{NGP}}\sin\delta + \cos\delta_{\mathrm{NGP}}\cos\delta\cos(\alpha-\alpha_{\mathrm{NGP}})\,,
    \end{equation}
    \begin{equation}
    \cos b\sin(l_{\mathrm{NCP}}-l) = \cos\delta\sin(\alpha - \alpha_{\mathrm{NGP}})\,,
    \end{equation}
    \begin{equation}
    \cos b\cos(l_{\mathrm{NCP}}-l) = \cos\delta_{\mathrm{NGP}}\sin\delta - \sin\delta_{\mathrm{NGP}}\cos\delta\cos(\alpha-
    \alpha_{\mathrm{NGP}})\,.
    \end{equation}

The angular position of SNe Ia are shown in Figure.~\ref{fig:
figure_l_b}.

\subsection{Distance modulus}
The analysis of SNe light curves (\citealt{Betoule14}) gives the
distance estimation $\mu_{\mathrm{obs}}$ as
    \begin{equation}
    \mu_{\mathrm{obs}} =  m_B^{*} - M_B +\alpha x_1 - \beta c\,,
    \end{equation}
where $m_B^{*}$ corresponds to the observed peak magnitude in
rest-frame $B$-band and $M_B$ is the absolute $B$-band magnitude.
$x_1$ and $c$ are SALT2 shape parameter and color correction
respectively. $\alpha$ and $\beta$ are nuisance parameters in the
distance estimate to be determined.

We correct the effects of host galaxy properties assuming that the
absolute magnitude is related to the host stellar mass
($M_{\mathrm{stellar}}$) by a simple step function
(\citealt{Betoule14})
    \begin{displaymath}
    M_B = \left\{ \begin{array}{ll}
    M_B^1 & \textrm{if $M_{\mathrm{stellar}}<10^{10}M_{\odot}$.}\\
    M_B^1 + \Delta_M & \textrm{otherwise.}
    \end{array} \right.
    \end{displaymath}
Here, we introduce the theoretical distance modulus $\mu$ as
    \begin{equation}
    \mu_{\mathrm{th}} = 5\lg(\frac{d_L}{\mathrm{Mpc}}) + 25\,.
    \end{equation}
\subsection{The Hubble diagram covariance matrix}
\cite{Betoule14} assemble a $3N_{\mathrm{SN}} \times
3N_{\mathrm{SN}} = 2220 \times 2220$ covariance matrix for the
light-curve parameters, which includes statistical and systematic
uncertainties.
    \begin{eqnarray}
    \lefteqn{\mathrm{C}_{\eta} = \mathrm{C}_{\mathrm{stat}} + (\mathrm{C}_{\mathrm{pecvel}} + \mathrm{C}_{\mathrm{nonIa}})_{\mathrm{C} 11} + {}} \nonumber\\
    && {} + (\mathrm{C}_{\mathrm{cal}} + \mathrm{C}_{\mathrm{model}} + \mathrm{C}_{\mathrm{bias}} + \mathrm{C}_{\mathrm{host}} + \mathrm{C}_{\mathrm{dust}})_{\mathrm{re-evaluated}}\,.
    \end{eqnarray}
$\mathrm{C}_{\mathrm{stat}}$ is obtained from error propagation of
light-curve fit uncertainties. The systematic uncertainties include
seven components, namely the calibration uncertainty
$\mathrm{C}_{\mathrm{cal}}$, the light-curve model uncertainty
$\mathrm{C}_{\mathrm{model}}$, the bias correction uncertainty
$\mathrm{C}_{\mathrm{bias}}$,  the mass step uncertainty
$\mathrm{C}_{\mathrm{host}}$, the peculiar velocity uncertainty
$\mathrm{C}_{\mathrm{pecvel}}$ and the non-Ia events uncertainty
$\mathrm{C}_{\mathrm{nonIa}}$.

The covariance matrix (\citealt{Betoule14}) of the vector of
distance modulus estimate $\mu_{\mathrm{obs}}$ is
    \begin{equation}
    \mathrm{C} = \mathrm{A}\mathrm{C}_{\eta}\mathrm{A}^{\dagger} + \mathrm{diag}\bigg(\frac{5\sigma_z}{z\ln 10}\bigg)^2 + \mathrm{diag}(\sigma_{\mathrm{lens}}^2) + \mathrm{diag}(\sigma_{\mathrm{coh}}^2)\, ,
    \end{equation}
where $\mathrm{A} = \mathrm{A}_0 + \alpha \mathrm{A}_1 -\beta
\mathrm{A}_2$ is a $2220\times740$ matrix with $(\mathrm{A}_k)_{i,j}
= \delta_{i, 3j+k}$ such that $\boldsymbol{\mu} =
\mathrm{A}\boldsymbol{\eta} - M_{B}$. Here, $\boldsymbol{\eta} =
\bigg((m_B^*, x_1, c)_1,...,(m_B^*, x_1, c)_{740}\bigg)$. The
$\sigma_z$, $\sigma_{\mathrm{lens}}$ and $\sigma_{\mathrm{coh}}$
account for the uncertainty in cosmological redshift due to peculiar
velocities, the variation of magnitudes caused by gravitational
lensing, and the intrinsic variation in SN magnitude not described
by the other terms, respectively. \cite{Betoule14} suggest
$c\sigma_{z} = 150\,\mathrm{km\,s}^{-1}$ and $\sigma_{\mathrm{lens}}
= 0.055\times z$, while values of $\sigma_{\mathrm{coh}}$ are listed
in Table \ref{tab: sigma_coh}.
    \begin{table}
        \centering
        \caption{Values of $\sigma_{\mathrm{coh}}$ used in the cosmological fits for different samples.}
        \label{tab: sigma_coh}
        \begin{tabular}{ccccc} % four columns, alignment for each
            \hline
            $\boldsymbol{\mathrm{Sample}}$ & $\mathrm{low}-z$ & SDSS-II & SNLS & HST\\
            \hline
            $\boldsymbol{\sigma_{\mathrm{coh}}}$ & 0.134 & 0.108 & 0.080 & 0.100\\
            \hline
        \end{tabular}
    \end{table}
\subsection{Markov Chain Monte Carlo fitting}
We employ the emcee to carry out parameters fitting
(\citealt{Goodman10}; \citealt{Foreman-Mackey13}). The emcee is a
python module that employs the affine-invariant ensemble sampler for
Markov Chain Monte Carlo (MCMC) algorithm. In this package, many
samplers (called walkers) run in parallel and periodically exchange
states to more efficiently sample the parameter space. The Maximum
Likelihood Estimation (MLE) is applied to MCMC algorithm. The
likelihood $L$ is sum of many normal distributions
\begin{equation}
L =
\prod_{i=1}^{740}\frac{1}{\sqrt{2\pi}\sigma_i}\exp\bigg[\frac{-(\mu_{\mathrm{obs,i}}-\mu_{\mathrm{th,i}})^2}{2\sigma_i^2}\bigg]\,,
\end{equation}
where
\begin{equation}
\sigma_i = \mathrm{C}_{ii}\,.
\end{equation}
Then the log-likelihood is
\begin{equation}
\ln L =
-\frac{1}{2}\sum_{i=1}^{740}\bigg[\frac{(\mu_{\mathrm{obs,i}}-\mu_{\mathrm{th,i}})^2}{\sigma_i^2}+\ln
(\sigma_i^2) + \ln (2\pi)\bigg]\,.
\end{equation}
The observational distance modulus $\mu_{\mathrm{obs}}$
depends on light-curve parameters \{$\alpha, \beta, M_B^1,
\Delta_M$\}, while the theoretical distance modulus
$\mu_{\mathrm{th}}$ depends on cosmological parameters \{$H_0$,
$\Omega_\mathrm{m}, \Sigma_0, w, \delta, l_A, b_A$\} in ellipsoidal
universe model. The priors of parameters are listed in Table
\ref{tab: priors}.

\cite{Campanelli11c} studied an anisotropic Bianchi type I
cosmological model using Union2 compilation, in which
$\mu_{\mathrm{obs}}$ and $\sigma$ are derived in the $\Lambda$CDM
model (\citealt{Amanullah10}). Therefore, the light-curve parameters
\{$\alpha, \beta, M_B^1, \Delta_M$\} are fixed. However, the
light-curve parameters vary with different cosmological model. It's
unreasonable to fit just cosmological parameters. Different
from \cite{Campanelli11c}, we fit four light-curve parameters and
seven cosmological parameters \{$H_0, \Omega_\mathrm{m}, \Sigma_0,
w, \delta, l_A, b_A$\} simultaneously. So the likelihood function
depends on eleven free parameters. The derived results are
model-independent.

As a comparison, we carried out another MCMC fitting which
constrained parameters \{$\alpha, \beta, M_B^1, \Delta_M, H_0,
\Omega_\mathrm{m}, w$\} in a flat $w$CDM cosmology with an arbitrary
equation of state $w$.

    \begin{table}
        \centering
        \caption{The priors of free parameters in MCMC fitting.}
        \label{tab: priors}
        \begin{tabular}{cccc}
            \hline
            Parameter & Prior & Parameter & Prior\\
            \hline
            $\alpha$ & [0.1, 0.2] & $\Sigma_0$ & [-0.5, 0.5]\\
            $\beta$ & [2.0, 4.0] & $w$ & [-2.0, 0.0]\\
            $M_B^1$ & [-19.3, -18.6] & $\delta$ & [-0.5, 0.5]\\
            $\Delta_M$ & [-0.1, 0.0] & $l_A$ & [0, $\pi$]\\
            $H_0$ & [60, 80] & $b_A$ & [$-\frac{\pi}{2}$, $\frac{\pi}{2}$]\\
            $\Omega_{\mathrm{m}}$ & [0.0, 0.5] & &\\
            \hline
        \end{tabular}
    \end{table}

\section{Results}
\label{sec: result} 
\begin{figure*}
        \centering
        \includegraphics[width=7in]{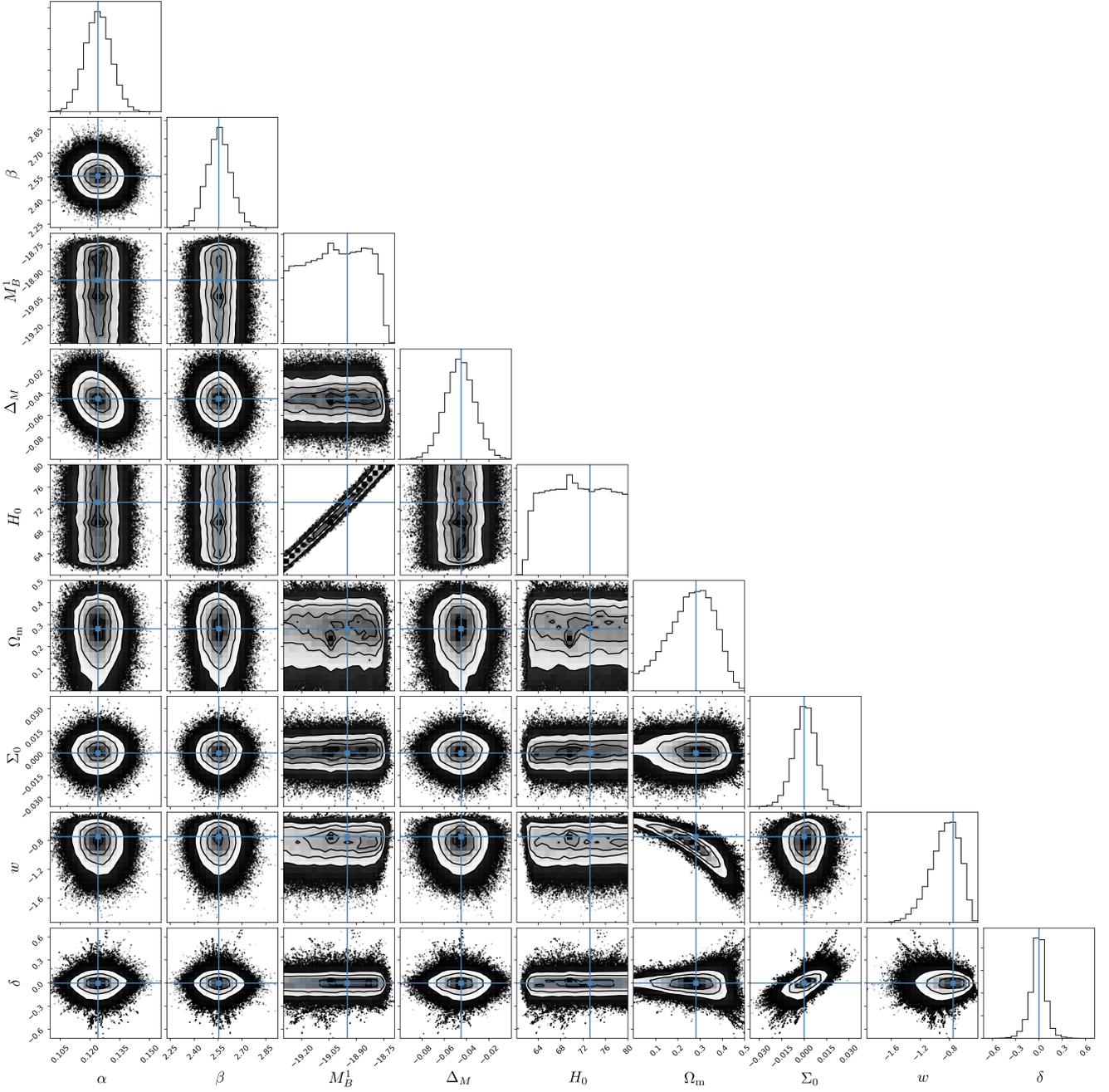}
        \caption{Confidence contours ($1\sigma$, $2\sigma$ and $3\sigma$) and marginalized likelihood distributions for the parameters ($\alpha, \beta, M_B^1, \Delta_M, H_0, \Omega_\mathrm{m}, \Sigma_0, w, \delta$) in
        ellipsoidal universe model. True values (shown as dots) of \{$\alpha, \beta, M_B^1, \Delta_M, \Omega_\mathrm{m}, w$\}
are taken from $w$CDM fitting results as comparison.}
        \label{fig: mcmc_ellipsoidal}
    \end{figure*}

The confidence contours ($1\sigma$,
$2\sigma$ and $3\sigma$) and marginalized likelihood distribution
functions for the parameters ($\alpha, \beta, M_B^1, \Delta_M, H_0,
\Omega_\mathrm{m}, \Sigma_0, w, \delta$) from MCMC fittings are
shown in Figure \ref{fig: mcmc_ellipsoidal}. Table \ref{tab:
parameters} lists the best-fitting value and the $1\sigma$,
$2\sigma$ and $3\sigma$ confidence level intervals. Since
$M_B^1$ entirely degenerates with $H_0$, their values can not be
constrained well simultaneously. If we take Hubble constant $H_0 =
73.24\,~\mathrm{km\,s}^{-1} \mathrm{Mpc}^{-1}$ \citep{Riess16}, the
corresponding $M_B^1$ is -18.95.

The constraints on the anisotropy parameters $\delta$ and $\Sigma_0$
from JLA sample are
    \begin{equation}
    -0.101<\delta<0.071 \quad(1\sigma)\,,
    \end{equation}
    and
    \begin{equation}
    -0.007<\Sigma_0<0.008 \quad(1\sigma)\,.
    \end{equation}
Compared with $-0.16<\delta<0.12$ and $-0.012<\Sigma_0<0.012$
($1\sigma$) from Union2 sample \citep{Campanelli11c}, we can see
that our result is more tight. The result implies that there is no
evidence in favor of either geometric anisotropy ($\Sigma_0\ne 0$)
or dark energy anisotropy ($\delta\ne 0$).

\begin{table*}
        \centering
        \caption{Best-fitting values, and the $1\sigma$, $2\sigma$ and $3\sigma$ confidence level intervals derived from the JLA sample.}
        \label{tab: parameters}
        \begin{tabular}{cccccccc} % four columns, alignment for each
            \hline
            \qquad & $\alpha$ & $\beta$ & $\Delta_M$ & $\Omega_{\mathrm{m}}$ & $\Sigma_0$ & $w$ & $\delta$\\
            \hline
            BF & 0.124 & 2.554 & -0.045 & 0.314 & 0.001 & -0.774 & -0.008\\
            $1\sigma$ & [0.118, 0.131] & [2.475, 2.622] & [-0.059, -0.032] & [0.173, 0.383] & [-0.007, 0.008] & [-1.028, -0.633] & [-0.101, 0.071]\\
            $2\sigma$ & [0.111, 0.137] & [2.406, 2.701] & [-0.072, -0.020] & [0.046, 0.443] & [-0.015, 0.015] & [-1.296, -0.535] & [-0.236, 0.185]\\
            $3\sigma$ & [0.105, 0.144] & [2.335, 2.777] & [-0.086, -0.007] & [0.003, 0.488] & [-0.024, 0.023] & [-1.527, -0.468] & [-0.451, 0.388]\\
            \hline
        \end{tabular}
    \end{table*}

\begin{figure}
        \centering
        \includegraphics[width=2.5in]{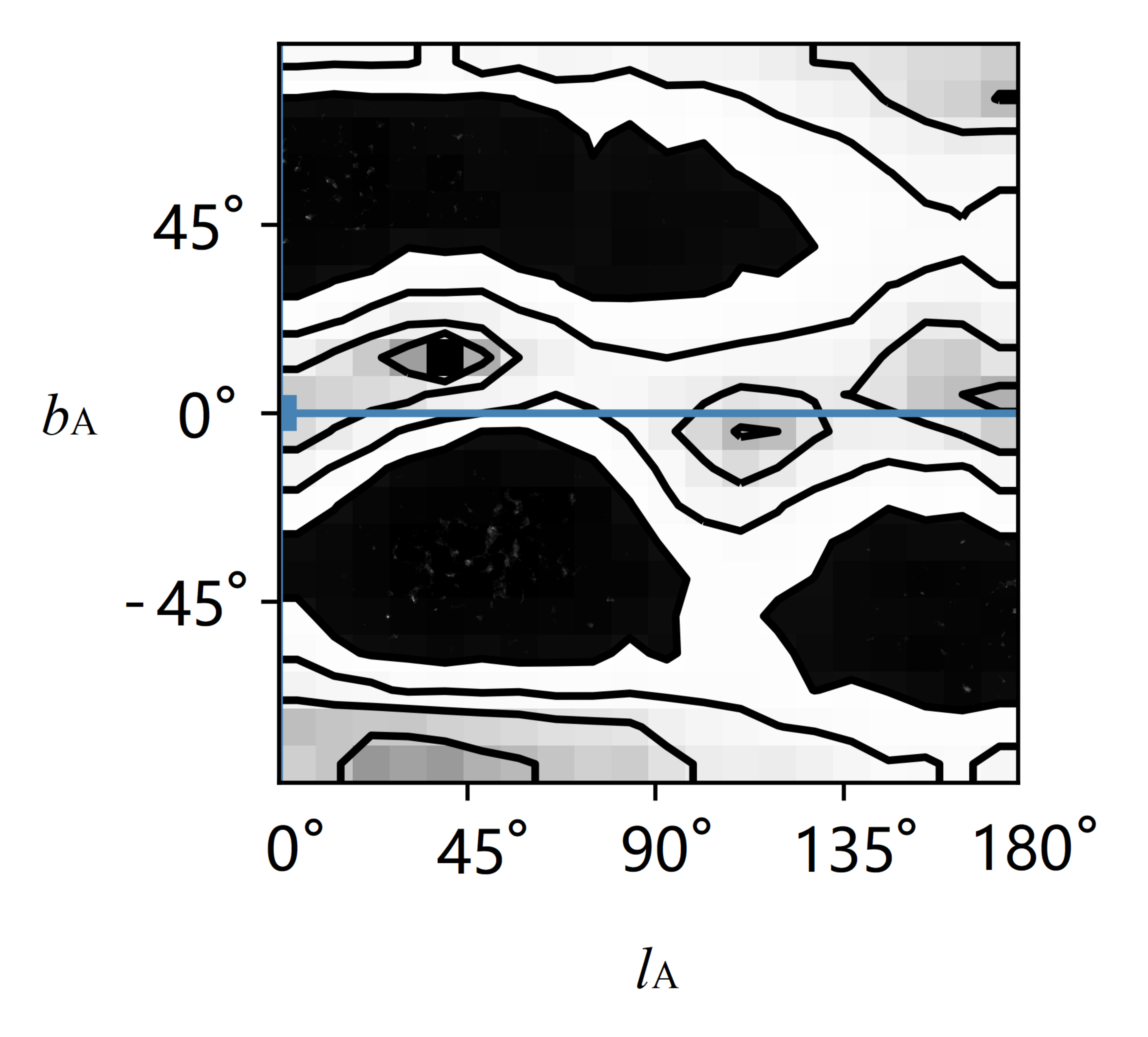}
        \caption{Confidence level contours of preferred direction in the galactic coordinate system.}
        \label{fig: lA_bA}
    \end{figure}

The direction cosine of anisotropy axis is
    \begin{equation}
    \hat{n}_A = (\cos b_A\cos l_A, \cos b_A\sin l_A, \sin b_A)\,.
    \end{equation}
The opposite direction cosine $\hat{n}_{A}'$ is then
    \begin{equation}
    \hat{n}_{A}' = (-\cos b_A\cos l_A, -\cos b_A\sin l_A, -\sin b_A)\,.
    \end{equation}
corresponding to an opposite direction ($l_A\pm\pi, -b_A$) of the
same axis. Therefore, we just use the parameter space $[0, \pi]$ for
$l_A$. Figure \ref{fig: lA_bA} gives confidence contours of
preferred direction in the galactic coordinate system. Since the
distance modulus depends weakly on both the angular position of SNe
Ia, $\hat{n}$, and the direction of the symmetry axis $\hat{n}_A$
for small values of the shear $\Sigma_0$, the SNe Ia data are not
able to constrain the anisotropy parameters ($\Sigma_0$ and
$\delta$) and the preferred direction defined by the anisotropy
itself at the same time.

The anti-correlation between $w$ and $\Omega_\mathrm{m}$ comes from
the dependence of luminosity distance on $w$ and
$\Omega_{\mathrm{m}}$
    \begin{equation}
    \bar{H} = \sqrt{\Omega_{\mathrm{m}}A^{-3} + \Omega_{\mathrm{DE}}A^{-3(1+w)}}\, .
    \end{equation}
Meanwhile, the correlation between $\delta$ and $\Sigma_0$
(\citealt{Campanelli11c}) is given by
    \begin{equation}
    \Sigma(A) = \frac{\Sigma_0 + (E-E_0)\delta}{A^3\bar{H}}\,,
    \end{equation}
which shows effects of energy distribution on the spatial curvature
of the universe.
Meanwhile, we find the best-fitting value for matter density parameter
    $\Omega_{\mathrm{m}} = 0.314_{-0.131}^{+0.069}\quad (1\sigma)\,,$
    and dark energy equation of state
    $w = -0.774_{-0.254}^{+0.141}\quad (1\sigma)\,. $

\begin{figure*}
        \centering
        \includegraphics[width=5.4in]{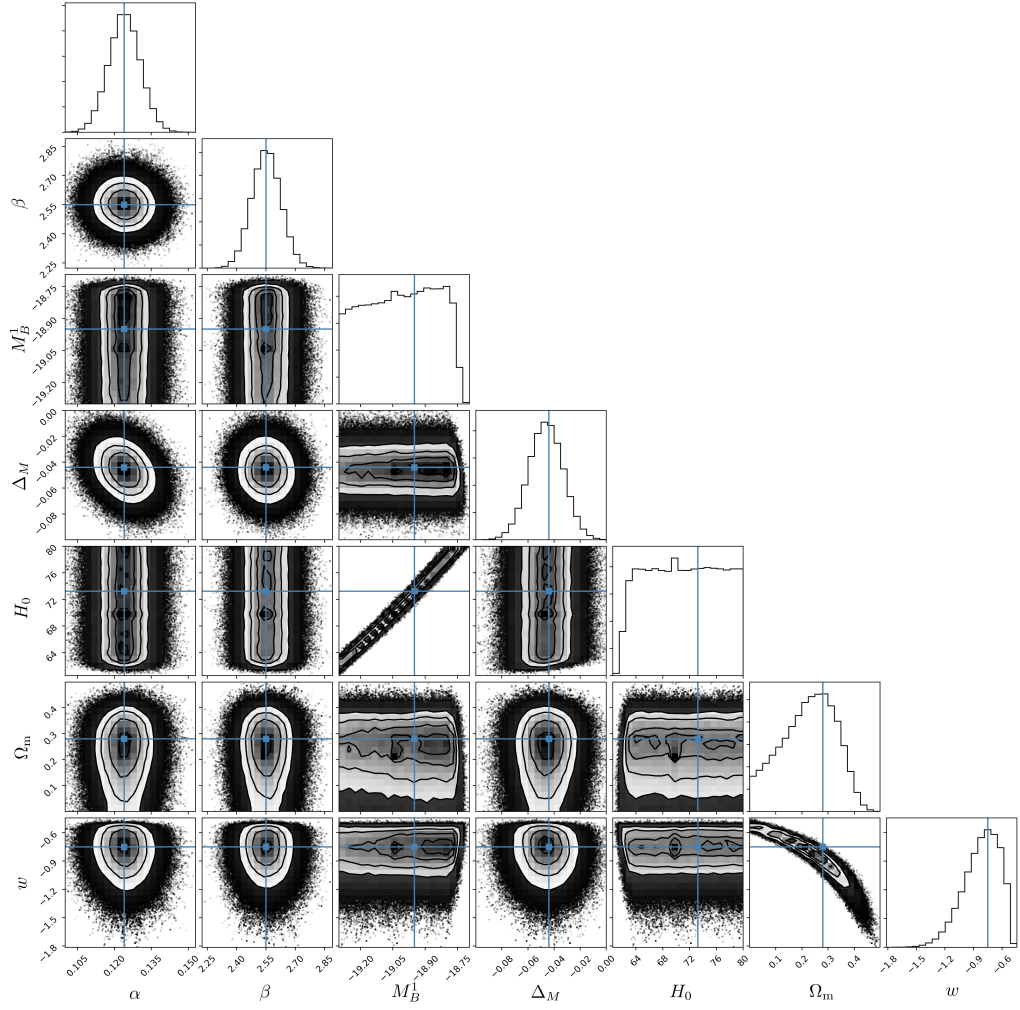}
        \caption{Confidence contours ($1\sigma$, $2\sigma$ and $3\sigma$) and marginalized likelihood distribution functions
        for the parameters ($\alpha, \beta, M_B^1, \Delta_M$, $H_0$, $\Omega_\mathrm{m}, w$) in the $w$CDM model.}
        \label{fig: mcmc_wCDM}
    \end{figure*}

\begin{figure*}
        \centering
        \includegraphics[width=4.5in]{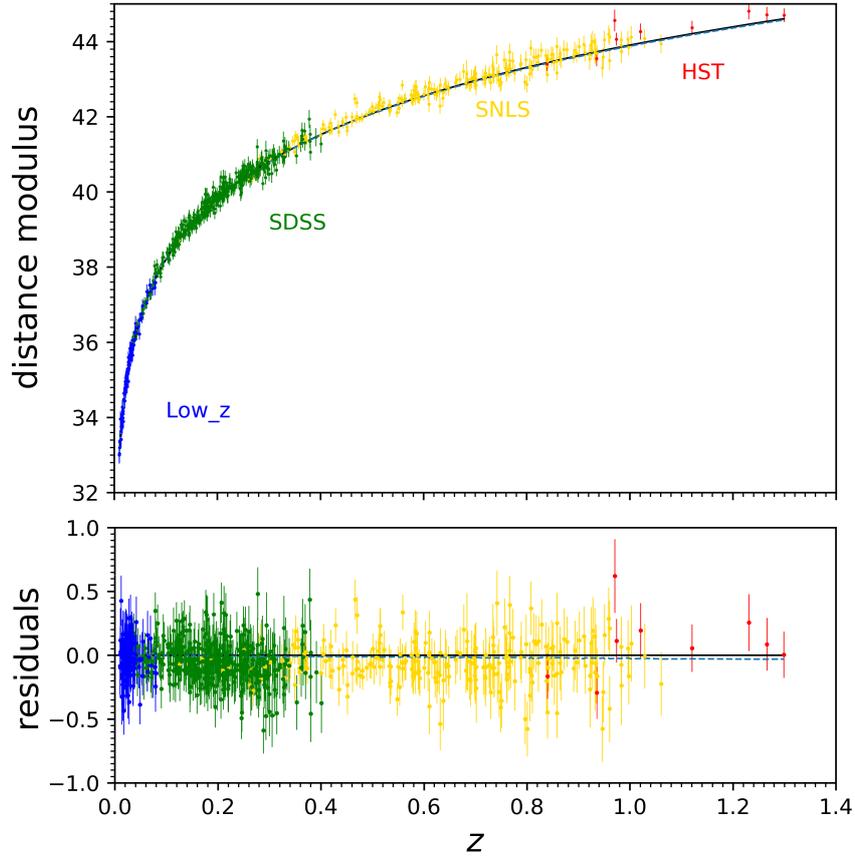}
        \caption{$Upper panel$. Hubble diagram for the 740 SNe Ia in the JLA compilation for different cosmological models:
        best-fitting ellipsoidal model $(\Omega_{\mathrm{m}}, \Sigma_0, w, \delta)\simeq (0.314, 0.001, -0.774, -0.008)$
        (blue dashed line), and  $w$CDM model ($\Omega_{\mathrm{m}}, w)\simeq (0.281, -0.750)$ (black line). $Lower panel$.
        Residuals (distance modulus minus distance modulus for the $w$CDM model) for the same models in the upper panel.}
        \label{fig: Campanelli_Hubble_diagram}
    \end{figure*}

The best-fitting results of the flat $w$CDM model are
displayed in Figure \ref{fig: mcmc_wCDM} for comparison. We find
that best-fitting values of parameters in ellipsoidal model are
quite similar to those in the $w$CDM model (see Table \ref{tab:
models}). Considering that the best-fitting values of $\Sigma_0$ and
$\delta$ are nearly zero, and $w$CDM is the limiting case of
ellipsoidal universe model, this result can be expected. Figure
\ref{fig: Campanelli_Hubble_diagram} shows the Hubble diagram for
JLA sample in two different cosmological models. One is the
best-fitting model with $(\Omega_{\mathrm{m}}, \Sigma_0, w, \delta)=
(0.314, 0.001, -0.774, -0.008)$ (blue dashed line). The other is the
flat $w$CDM model with $\Omega_{\mathrm{m}}= 0.281$, $w= -0.750$
(black line). In the lower panel, we show the corresponding distance
modulus fitting residuals (distance modulus minus the best-fitting
$w$CDM distance modulus) as a function of redshift. The comparison
between the best-fitting values of cosmological parameters from
Union2 (\citealt{Campanelli11c}) and JLA sample is shown in Table
\ref{tab: comparison}. The constraints are more tight than those of
\cite{Campanelli11c}.
    \begin{table}
        \centering
        \caption{Comparisons between best-fitting values in $w$CDM model and ellipsoidal universe model.}
        \label{tab: models}
        \begin{tabular}{ccccccc} % four columns, alignment for each
            \hline
            & $\alpha$ & $\beta$ & $M_{B}^1$ & $\Delta_{M}$ & $\Omega_{\mathrm{m}} $ & $w$\\
            \hline
            $w$CDM & 0.124 & 2.554 & -18.95 & -0.045 & 0.281 & -0.750\\
            \hline
            Ellipsoidal & 0.124 & 2.554 & -18.95 & -0.045 & 0.314 & -0.774\\
            \hline
        \end{tabular}
    \end{table}

    \begin{table}
        \centering
        \caption{Comparisons between best-fitting values of cosmological parameters from Union2 and JLA sample for ellipsoidal universe model.}
        \label{tab: comparison}
        \begin{tabular}{ccccc} % four columns, alignment for each
            \hline
            & $\Sigma_0$ & $\delta$ & $\Omega_{\mathrm{m}}$ & $w$\\
            \hline
            $\mathrm{Union2}$ & -0.004 & -0.050 & 0.37 & -1.32\\
            \hline
            $\mathrm{JLA}$ & 0.001 & -0.008 & 0.314 & -0.774\\
            \hline
        \end{tabular}
    \end{table}

\section{Conclusions}
In this paper, we study the ellipsoidal universe model with
plane-symmetric metric and constrain the anisotropy level of cosmic
geometry and dark energy fluids. By analyzing the magnitude-redshift
data of 740 SNe Ia in the JLA sample, we find a more tight
constraint on cosmic shear $\Sigma_0$ and skewness $\delta$. The
best constraints are
    \begin{equation}
    -0.007<\Sigma_0<0.008 \quad(1\sigma)\,, \nonumber
    \end{equation}
    and
    \begin{equation}
    -0.101<\delta<0.071 \quad(1\sigma)\,. \nonumber
    \end{equation}
In conclusion, the fitting results favor an isotropic universe
without a preferred direction at present time. With the progress of
astronomical observations such SNe Ia and CMB, we will have better
constraints on universe anisotropy in the near future. The question
that whether the cosmic anomalies such as dark energy dipole, fine
structure constant dipole or dark flow have same physical origin
remains to be answered. Cosmological principle is so vital to modern
cosmology that much more effort should be made to verify the
fundamental postulation.
    \label{sec:conclusions}

    %\section*{Acknowledgements}

\section*{Acknowledgements}
We thank the anonymous referee for constructive comments. This work
is supported by the National Basic Research Program of China (973
Program, grant No. 2014CB845800) and the National Natural Science
Foundation of China (grants 11422325 and 11373022), and the
Excellent Youth Foundation of Jiangsu Province (BK20140016).
    %%%%%%%%%%%%%%%%%%%%%%%%%%%%%%%%%%%%%%%%%%%%%%%%%%



    %\bibliographystyle{mnras}
\newpage


    %%%%%%%%%%%%%%%%%%%%%%%%%%%%%%%%%%%%%%%%%%%%%%%%%%
%\newpage
    \suppressfloats
    \begin{figure*}
        \includegraphics[width=6in]{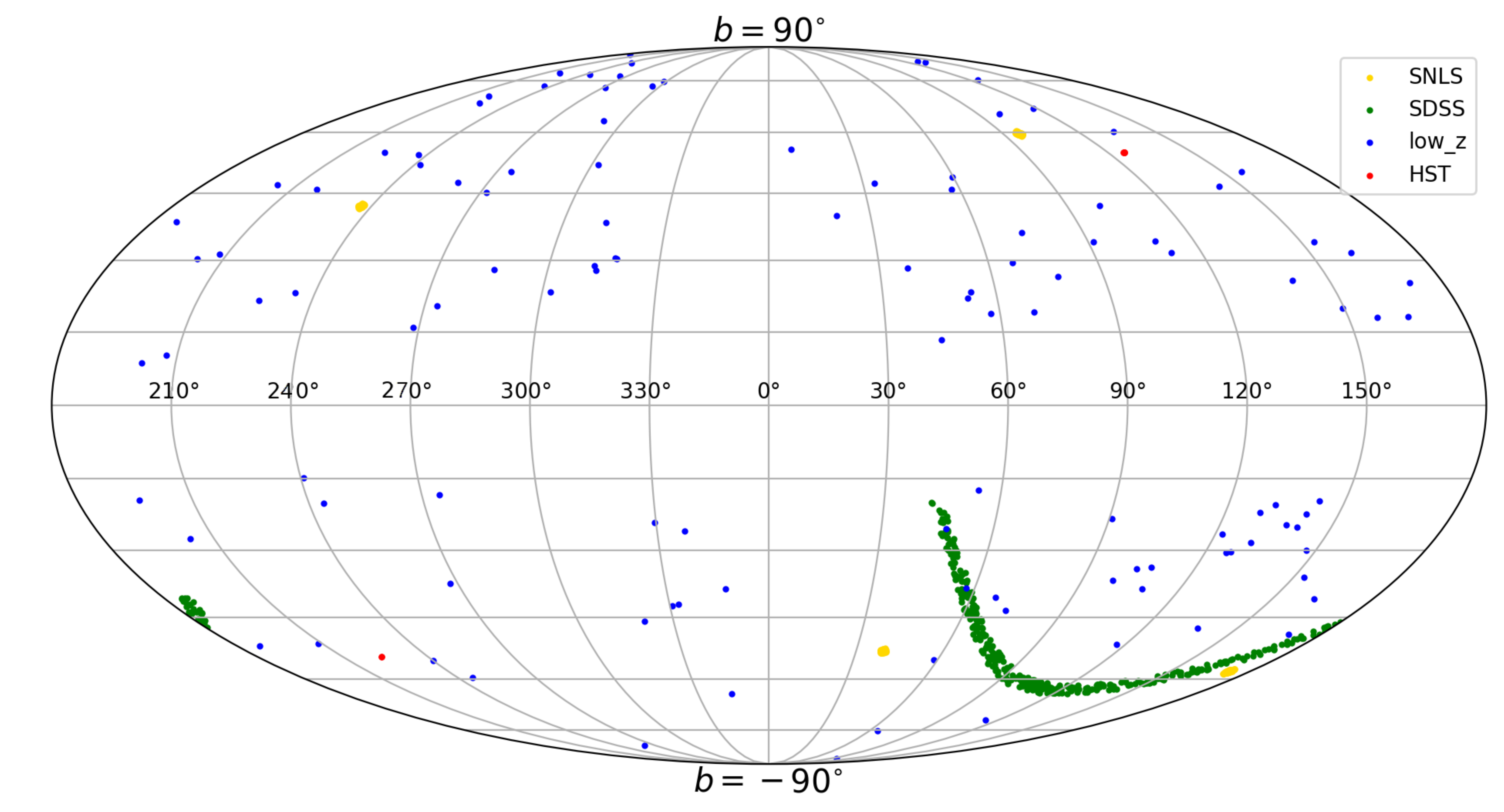}
        \caption{Angular positions of SNe Ia in the JLA sample. $b$ ($-90^{\circ}\le b\le 90^{\circ}$) is the galactic latitude and $l$ ($0^{\circ}\le l< 360^{\circ}$) is the galactic longitude. Supernovae from four subsets are marked with different colors.}
        \label{fig: figure_l_b}
    \end{figure*}

    %%%%%%%%%%%%%%%%%%%%%%%%%%%%%%%%%%%%%%%%%%%%%%%%%%
    % Don't change these lines
    \bsp    % typesetting comment
    \label{lastpage}
\end{document}